\documentclass[12pt]{article}
\newcommand{\lesssim}{\:\mbox{\raisebox{-3pt}{$\stackrel%
{\displaystyle <}{\sim}$}}\:}
\newcommand{\mms}{\Delta m^2_\odot}
\newcommand{\mma}{\Delta m^2_\mathrm{atm}}
\newcommand{\lll}{\ll\hskip-5.5pt <}
\begin{document}
\title{\normalsize \hfill UWThPh-2001-41 \\[0.8cm] \LARGE
Softly broken lepton numbers: \\
an approach to maximal neutrino mixing\thanks{Talk presented by
W.\ Grimus at the
XXV International School of Theoretical Physics, Ustro\'n, Poland,
September 10--16, 2001}}
\author{Walter Grimus\thanks{E-mail: grimus@doppler.thp.univie.ac.at} \\
\setcounter{footnote}{3}
\small Institut f\"ur Theoretische Physik, Universit\"at Wien \\
\small Boltzmanngasse 5, A--1090 Wien, Austria \\*[3.6mm]
Lu\'{\i}s Lavoura\thanks{E-mail: balio@cfif.ist.utl.pt} \\
\small Universidade T\'ecnica de Lisboa \\
\small Centro de F\'\i sica das Interac\c c\~oes Fundamentais \\
\small Instituto Superior T\'ecnico, P--1049-001 Lisboa, Portugal \\*[4.6mm] }

\vspace{-0.6cm}

\date{3 October 2001}

\maketitle

\vspace{-0.5cm}

\begin{abstract}
We discuss models where
the $U(1)$ symmetries of lepton numbers
are responsible for maximal neutrino mixing. We pay particular
attention to an extension of the Standard Model
with three right-handed neutrino
singlets in which
we require that the three lepton numbers $L_e$, $L_\mu$, and $L_\tau$
be separately conserved in the Yukawa couplings, but assume that they
are \emph{softly} broken by the Majorana mass matrix $M_R$
of the neutrino singlets.
In this framework,
where lepton-number breaking occurs at a scale
much higher than the electroweak scale,
deviations from family lepton number conservation are
calculable, i.e., finite,
and lepton mixing stems exclusively from $M_R$.
We show that in this framework either maximal atmospheric neutrino
mixing or maximal solar neutrino mixing or both can be imposed
by invoking symmetries. In this way
those maximal mixings
are stable against radiative corrections.
The model which achieves maximal (or nearly maximal) solar neutrino
mixing  assumes that there are two different scales in $M_R$
and that the lepton number $\bar L = L_e - L_\mu - L_\tau$ is
conserved in between them. We work out the difference between this
model and the conventional scenario where (approximate) $\bar L$
invariance
is imposed directly on the mass matrix of the \emph{light} neutrinos.
\end{abstract}

\newpage

\section{Neutrino masses, mixing and oscillations}

The impressive results of the atmospheric \cite{atm} and solar
\cite{solar} neutrino experiments have a natural explanation in terms
of neutrino oscillations
\cite{pontecorvo,review}.
With respect to solar neutrinos,
this point of view was further strengthened---and the possibility
of an astrophysical solution \cite{bahcall-SSM} further weakened---by
the first results of the SNO experiment \cite{SNO}.
In this context, the large mixing angle (LMA) MSW solution
of the solar neutrino deficit \cite{MSW},
with a solar mass-squared difference $\mms \sim 5 \times 10^{-5}$ eV$^2$,
is emerging as the favoured scenario
\cite{marfatia,fogli,bahcall,goswami,krastev,garzelli};
whereas the small
mixing (SMA) MSW solution seems to fade. The LOW solution, with
$\mms \sim 10^{-7}$ eV$^2$, also gives a reasonable fit to the solar
neutrino data. For the LMA MSW solution the solar mixing angle
$\theta$ is large, but $\theta = 45^\circ$ is not allowed at 99\% CL;
for the LOW solution maximal mixing gives a somewhat
better fit \cite{krastev}. In the case of atmospheric neutrinos,
the fit to the deficit of the muon-neutrino flux
leads to the best-fit values
$\mma = 2.5 \times 10^{-3}$ eV$^2$ and $\psi = 45^\circ$ \cite{nishikawa},
with the atmospheric mixing angle $\psi$
sitting squarely on
maximal mixing.

In the following we shall consider exclusively neutrino mixing and
oscillations among the known three active neutrinos. We shall not
take into consideration the result of the LSND experiment \cite{LSND}
which---when interpreted in terms of neutrino oscillations---leads to a
mass-squared difference
$\Delta m^2_\mathrm{LSND} \sim 1$ eV$^2$.

Neutrino mixing relates the left-handed neutrino flavour fields with
the neutrino fields with definite mass $m_j$ via
\begin{equation}\label{mixing}
\nu_{\alpha L} = \sum_j U_{\alpha j} \nu_{jL}
\quad (\alpha = e, \mu, \tau) \,,
\end{equation}
where $U$ is a unitary $3 \times 3$ matrix.
Plugging Eq.~(\ref{mixing}) into the charged-current Lagrangian
we obtain
\begin{equation}\label{CC}
\mathcal{L}_{CC} =
\frac{g}{\sqrt{2}}\,
W^-_\rho\!
\sum_{\alpha = e,\mu,\tau}
\bar{\ell}_\alpha \gamma^\rho \,
\sum_j U_{\alpha j} \nu_{jL} + \mathrm{h.c.}
\end{equation}
Neutrino flavour eigenstates,
which are coherent superpositions of the neutrino mass eigenstates,
are produced and detected via charged-current interactions
such that the neutrino flavour is defined via the
charged lepton associated with the production or detection process.
It is crucial in this context that $E^2 \gg m_j^2$, where $E$
is the neutrino energy, holds in any realistic experiment.
The general formula for the neutrino survival ($\alpha = \beta$) and
transition ($\alpha \neq \beta$) probabilities is given by
\cite{pontecorvo}
\begin{equation}\label{prob}
P_{\nu_\alpha \to \nu_\beta} (L/E) =
\left| \sum_j U_{\beta j} U_{\alpha j}^* \exp\, (-i m_j^2 L/2E)
\right|^2 \,,
\end{equation}
where $L$ denotes the distance in between the neutrino source and
neutrino detection points. These probabilities depend on the ratio
$L/E$ and on the neutrino
mass-squared differences.\footnote{For $n$ neutrinos there are $n-1$
independent mass-squared differences.}

In the following
we shall use the convention $m_1 < m_2$
for the ordering of the neutrino masses.
Experimentally we know that $\mms \ll \mma$,
hence we take $\mms = m_2^2 - m_1^2$ to be much smaller
than $\mma = \left| m_3^2 - m_1^2 \right|$.
We have to distinguish between the two mass spectra
$m_3 > m_1 \simeq m_2$ and $m_3 < m_1 \simeq m_2$.
The atmospheric neutrino mixing angle will be denoted by $\psi$
and the solar mixing angle by $\theta$.

It is an interesting fact that there are no indications in favour of
electron-neutrino oscillations, neither in long baseline \cite{CHOOZ} nor
in atmospheric neutrino experiments \cite{atm,nishikawa}. This leads to
the conclusion that $U_{e3}$ is small. A three-neutrino fit \cite{GG-valle}
to the CHOOZ,
solar, and atmospheric neutrino data (before SNO) leads to the upper bound
$|U_{e3}|^2 \lesssim 0.04$.
It is instructive to explore the limit
$U_{e3} \to 0$, where decoupling of solar and atmospheric neutrino
oscillations takes place \cite{giunti}. In this limit, the mixing
matrix factorizes:
\begin{equation}\label{U0}
U' = U_\mathrm{atm} U_\odot =
\left( \begin{array}{ccc}
1 & 0 & 0 \\
0 & \cos\psi & \sin\psi \\
0 & -\sin\psi & \cos\psi
\end{array} \right)
\left( \begin{array}{ccc}
\cos\theta & \sin\theta & 0 \\
-\sin\theta & \cos\theta & 0 \\ 0 & 0 & 1
\end{array} \right)
\end{equation}
or
\begin{equation}\label{U00}
U = \mathrm{diag}\, (1,-1,1)\, U' =
\left( \begin{array}{ccc}
\cos\theta & \sin\theta & 0 \\
\cos\psi \sin\theta & -\cos\psi \cos\theta & -\sin\psi \\
\sin\psi \sin\theta & -\sin\psi \cos\theta & \cos\psi
\end{array} \right) \,.
\end{equation}
In Eqs.~(\ref{U0}) and (\ref{U00}) all unphysical phases have been
removed and a special phase convention, which will be used in the
following, has been chosen in Eq.~(\ref{U00}).
In this picture one can show that the following transitions are realized:
\begin{equation}
U_{e3} = 0 \; \Rightarrow \; \left\{
\begin{array}{rl}
\mbox{solar $\nu$s:} &
| \nu_e \rangle \to
\cos\psi\, |\nu_\mu \rangle - \sin\psi\, | \nu_\tau \rangle \,, \\
\mbox{atmospheric $\nu$s:} &
| \nu_\mu \rangle \to | \nu_\tau \rangle \quad \mbox{and} \quad
| \bar\nu_\mu \rangle \to | \bar\nu_\tau \rangle \,.
\end{array} \right.
\end{equation}

Let us summarize the picture which emerges from the comparison of
three-neutrino oscillations with the solar and atmospheric
neutrino data:
\begin{enumerate}
\item
$\mms/\mma \ll 1$; \label{ratio}
\item
$|U_{e3}|^2 \ll 1$;
\item
$\theta$ large, but not $45^\circ$;
\item
$\psi \simeq 45^\circ$.
\end{enumerate}
In point \ref{ratio}, the ratio $\mms / \mma$ is of order
$10^{-2}$ for the LMA MSW solution,
$10^{-5}$--$10^{-4}$ for the LOW solution.

To these four points one could add the requirement that the neutrino
masses are much smaller than the charged-lepton masses. One popular
way of satisfying this requirement is given by the seesaw mechanism
\cite{seesaw}, which will be adopted in this paper. Concerning points
1--4, it seems difficult to
identify plausible mechanisms to ``explain'' them, despite huge efforts
in this direction. In particular, the
feature that atmospheric mixing is maximal whereas solar mixing is large
\emph{but not maximal} represents a formidable task for model
building \cite{dorsner}.

In the present paper, which is based on Ref.~\cite{grimus},
we discuss the effects of softly broken lepton numbers on the question
of maximal or large neutrino mixing. We aim at obtaining those mixings
from symmetries, in particular from the $U(1)$ invariances associated with
lepton numbers.
In that way,
when we are able to enforce (at least some of)
the features enumerated above,
our results will be stable against radiative corrections.
We want to stress that this is not the case
if ``textures'' or \emph{Ans\"atze} are used for that end.

\section{$L_e - L_\mu - L_\tau$ invariance}
\label{Lbarinv}

In this section we want to discuss the well-known \cite{barbieri}
example of
$\bar L \equiv L_e - L_\mu - L_\tau$ invariance of the 
\emph{light-neutrino mass matrix}
and its connection with maximal solar mixing.
We do not consider
heavy right-handed singlets $\nu_R$ in this section.

We need the mass terms for the neutrinos
\begin{equation}\label{Lmnu}
\mathcal{L}^{(\nu)}_m = \frac{1}{2}
\nu_L^T C^{-1} \mathcal{M}_\nu \nu_L + \mbox{h.c.}
\end{equation}
and for the charged leptons
\begin{equation}\label{Lml}
\mathcal{L}^{(\ell)}_m = - \bar\ell_R M_\ell \ell_L + \mbox{h.c.}
\end{equation}
Invariance of the Lagrangian under $\bar L$ is an abbreviation for having
a symmetry group $U(1)$ in the following sense: the
three left-handed lepton doublets $D_\alpha$ and the right-handed singlets
$\ell_{\alpha R}$ are multiplied by 
$e^{i\gamma}$, $e^{-i\gamma}$, $e^{-i\gamma}$,
respectively, where $\gamma$ takes on all real numbers. As an obvious
consequence the mass matrices of Eqs.~(\ref{Lmnu}) and (\ref{Lml})
have the forms
\begin{equation}\label{MM}
\mathcal{M}_\nu = \left(
\begin{array}{ccc} 0 & p & q \\ p & 0 & 0 \\ q & 0 & 0 \end{array}
\right), \quad {\rm and} \quad
M_\ell = \left( \begin{array}{ccc}
M_{ee} & 0 & 0 \\ 0 & M_{\mu\mu} & M_{\mu\tau} \\
0 & M_{\tau\mu} & M_{\tau\tau} \end{array} \right) \,,
\end{equation}
respectively.
The bi-diagonalization of the $M_\ell$ of Eq.~(\ref{MM}) does not
change the form of the $\mathcal{M}_\nu$ of Eq.~(\ref{MM}). Therefore,
without loss of generality we may assume that $M_\ell$ is diagonal
and positive, and that $p$ and $q$ are real and positive.

The neutrino mass matrix of Eq.~(\ref{MM}) has the eigenvalues
\begin{equation}\label{lambda}
\lambda_1 = m_1 = m_0 \,, \;
\lambda_2 = -m_2 = -m_0 \,, \;
\lambda_3 = m_3 = 0\,,
\quad \mbox{with} \quad
m_0 = \sqrt{p^2+q^2} \,.
\end{equation}
Therefore, we obtain the mass-squared differences
\begin{equation}
\mma = m_0^2 \quad \mbox{and} \quad \mms = 0 \,.
\end{equation}
Evidently, some small breaking of $\bar L$ invariance, whether soft or
spontaneous, is necessary in order to achieve $\mms \neq 0$.

The neutrino mass matrix is diagonalized by
\begin{equation}\label{VMV}
V^T \mathcal{M}_\nu V = \mathrm{diag}\, (m_1,m_2,m_3) \,,
\end{equation}
where the $3 \times 3$ unitary matrix $V$ is determined as
\begin{equation}\label{VU}
V = U\, \mathrm{diag}\, (1,-i,1)
\quad \mbox{with} \quad
U = \left( \begin{array}{ccc}
1/\sqrt{2} & 1/\sqrt{2} & 0 \\
\cos\psi/\sqrt{2} & -\cos\psi/\sqrt{2} & -\sin\psi \\
\sin\psi/\sqrt{2} & -\sin\psi/\sqrt{2} &  \cos\psi
\end{array} \right) \,.
\end{equation}
The main point of this mixing matrix is that it has maximal
solar mixing, as a consequence of the assumption of $\bar L$
invariance. The atmospheric mixing angle, given by
\begin{equation}
\cos\psi = \frac{r}{\sqrt{1+r^2}} \,, \quad
\sin\psi = \frac{1}{\sqrt{1+r^2}}
\quad \mbox{with} \quad r = \frac{p}{q} \,,
\end{equation}
will be large---if we do not allow for any finetuning---but
not necessarily close to $45^\circ$. Let us estimate the amount of
\emph{tuning} of $r$ for bringing $\psi$ into agreement with the range
obtained from a fit to the atmospheric neutrino data \cite{nishikawa}:
\begin{equation}
\sin^2 2\psi
= \frac{4r^2}{(1+r^2)^2} > 0.89
\quad \mbox{(90\% CL)} \quad \Rightarrow \quad
0.7 \lesssim r \lesssim 1.4 \,.
\end{equation}
Thus, the present data allow a rather wide range for $r$.
In addition, one has to take into account that the $U$ of Eq.~(\ref{VU})
was obtained at a stage where $\bar L$ was still unbroken. Breaking of
the lepton number might move the solar angle $\theta$ a little away
from $45^\circ$, but no significant corrections to $\theta$ and $\psi$
are expected.

Let us for instance consider a specific example of soft $\bar L$ breaking,
where the $\mathcal{M}_\nu$ of Eq.~(\ref{MM}) changes to
\begin{equation}\label{softLbar}
\mathcal{M}_\nu = \left(
\begin{array}{ccc} a & p & q \\ p & 0 & 0 \\ q & 0 & 0 \end{array}
\right).
\end{equation}
(Models with this $\mathcal{M}_\nu$
have been described,
for instance,
in Ref.~\cite{softLbar} and in the references therein.)
It is easy to check that
\begin{equation}
\mma \simeq m_0^2 \quad \mbox{and} \quad \mms \simeq 2 m_0 a \,.
\end{equation}
Without loss of generality we have chosen $a > 0$. We can express $a$
and the solar mixing angle by the mass-squared differences:
\begin{equation}
a \simeq \frac{\mms}{2\sqrt{\mma}}
\quad \mbox{and} \quad
\sin^2 2\theta \simeq 1 - \frac{1}{16} \left( \frac{\mms}{\mma} \right)^2 \,.
\end{equation}
The first relation shows that $a \ll p, \, q$, whereas from the second
relation we read off that, for all practical purposes, the solar mixing
angle remains $45^\circ$ even after $\bar L$ breaking of the type
considered in Eq.~(\ref{softLbar}).

It is a generic feature of models with approximate $\bar L$ invariance
in the mass matrix of the light neutrinos that corrections to $\theta =
45^\circ$ are suppressed by
$\left( \mms/\mma \right)^2$ \cite{nir}.
This is a
drawback in view of the fit to present solar data,
where a solar mixing angle
very close to $45^\circ$ is disfavoured. In addition, in the
scheme under discussion atmospheric mixing comes out large but not
necessarily close to $45^\circ$.
There is no explanation for $\psi \simeq 45^\circ$
by using $\bar L$ invariance alone,
as in this section.

\section{A framework for imposing maximal atmospheric and/or maximal
solar mixing}
\label{frame}

In this section we shall introduce a framework which will allow us to
implement either maximal atmospheric neutrino mixing (Section
\ref{maxatm}) or (nearly) maximal solar neutrino mixing (Section
\ref{maxsol}) or both (Section \ref{bimax}). This framework is completely
different from the one discussed in the previous section.

We begin with an extension of the Standard Model based on the
following ingredients: 
\begin{itemize}
\item
The lepton sector of the Standard Model (SM) with three families,
plus three right-handed singlets $\nu_R$;
\item
the seesaw mechanism;
\item
$n_H$ scalar doublets.
\end{itemize}
The seesaw mechanism provides us with a reason why the neutrino masses
are much smaller than the charged-lepton masses. The number $n_H$ of
Higgs doublets is arbitrary in this section.

In order to fix the notation we recapitulate the seesaw mechanism. The
scalar doublets
and their vacuum expectation values (VEVs) are denoted by
\begin{equation}
\phi_j = \left( \begin{array}{c}
\varphi_j^+ \\ \varphi_j^0 \end{array} \right)
\quad \mbox{and} \quad
\left\langle 0 \left| \varphi_j^0 \right| 0 \right\rangle
= \frac{v_j}{\sqrt{2}}\, .
\end{equation}
The right-handed neutrino singlets have a Majorana mass term
\begin{equation} \label{LM}
\mathcal{L}_M = \frac{1}{2}\, \nu_R^T C^{-1} \! M_R^* \nu_R +
\mbox{h.c.}\,,
\end{equation}
where $M_R$ is symmetric.
The Yukawa Lagrangian of the leptons is given by
\begin{equation} \label{lyukawa}
\mathcal{L}_{\rm Y} = - \sum_{j=1}^{n_H} \left[ \bar \ell_R
\left( \begin{array}{cc} \varphi_j^-, & {\varphi_j^0}^\ast \end{array} \right)
\Gamma_j
+ \bar \nu_R
\left( \begin{array}{cc} \varphi_j^0, & - \varphi_j^+ \end{array} \right)
\Delta_j
\right]
\left( \begin{array}{c} \nu_L \\ \ell_L \end{array} \right)
+ \mbox{h.c.}
\end{equation}
From this Lagrangian, we derive the charged-lepton mass matrix $M_\ell$
and the Dirac neutrino mass matrix $M_D$ as
\begin{equation} \label{Ml}
M_\ell = \frac{1}{\sqrt{2}} \sum_j v_j^\ast \Gamma_j
\quad \mbox{and} \quad
M_D = \frac{1}{\sqrt{2}} \sum_j v_j \Delta_j\, ,
\end{equation}
respectively. The Dirac and Majorana mass terms for the neutrinos are
summarized in
\begin{equation}\label{majorana}
\frac{1}{2} \left( \begin{array}{cc} \nu_L^T, & \nu_L^{\prime T}
\end{array} \right) C^{-1} \mathcal{M}_{D+M}
\left( \begin{array}{c} \nu_L \\ \nu_L^\prime \end{array} \right)
+ \mbox{h.c.}
\quad \mbox{with} \quad
\mathcal{M}_{D+M} =
\left( \begin{array}{cc} 0 & M_D^T \\ M_D & M_R
\end{array} \right) \,,
\end{equation}
where $\nu_L' \equiv (\nu_R)^c = C \bar \nu_R^T$.
The seesaw mechanism gives an
effective mass term (\ref{Lmnu}) for the light left-handed neutrinos
with mass matrix \cite{seesaw}
\begin{equation} \label{seesaw}
\mathcal{M}_\nu = -M_D^T M_R^{-1} M_D \,.
\end{equation}
In general the mass matrix of the charged leptons will not be diagonal
and, therefore, has to be bi-diagonalized by two unitary matrices
$V_R^\ell$ and $V_L^\ell$:
\begin{equation}
{V_R^\ell}^\dagger M_\ell V_L^\ell =
\mathrm{diag}\, (m_e,m_\mu,m_\tau) \,.
\end{equation}
Then we arrive at the neutrino mixing matrix $U$ by
\begin{equation}
{V_L^\ell}^\dagger V = e^{i \hat \alpha} U e^{i \hat \beta} \,,
\end{equation}
where $V$ is given by Eq.~(\ref{VMV}),
while $e^{i\hat\alpha}$ and
$e^{i\hat\beta}$ are diagonal matrices of phase factors.
The phases
$\hat \alpha$ can be absorbed into the charged-lepton fields.
The phases $\hat \beta$, usually called Majorana phases, are irrelevant
for neutrino oscillations---although they play an important role in
processes like neutrinoless double-beta decay.

Schematically, in the seesaw mechanism the mass of a light Majorana
neutrino is obtained as $m_\nu \sim (m_\nu^D)^2/m_R$, where
$m_\nu^D$ is a typical scale in the Dirac neutrino mass matrix $M_D$
and $m_R$ a typical order of magnitude of the eigenvalues of
$\sqrt{M_R^* M_R}$. Identifying $m_\nu$ with $\sqrt{\mma} \simeq 0.05$
eV, and $m_\nu^D$ with either $m_\mu$ or $m_\tau$,
one obtains the well-known order of magnitude for the right-handed scale
$m_R \sim 10^8$--$10^{11}$ GeV.

Having discussed the conventional seesaw extension of the SM we
come to the main point which defines our framework:
\begin{center}
\fbox{\fbox{\parbox{10cm}{\textbf{ASSUMPTION:} \\
The lepton numbers $L_e$, $L_\mu$, and $L_\tau$ are \emph{conserved} in
the $\mathcal{L}_\mathrm{Y}$ of Eq.~(\ref{lyukawa}),
but they are \emph{softly broken} in
the $\mathcal{L}_M$ of Eq.~(\ref{LM}).
}}}
\end{center}
Note that the Majorana mass term of the right-handed neutrino
singlets, $\mathcal{L}_M$, has dimension 3; therefore, the breaking of
the family lepton numbers is indeed soft. As a consequence of our
assumption, the Yukawa coupling matrices $\Gamma_j$, $\Delta_j$ are
diagonal. Therefore,
\begin{itemize}
\item
$V_L^\ell = \mathbf{1}$;
\item
$M_D$ is diagonal;
\item
neutrino mixing stems exclusively from $M_R$;
\item
family lepton numbers $L_{e,\mu,\tau}$ and the total lepton number
$L = L_e+L_\mu+L_\tau$ are broken at the heavy scale $m_R$, which is
much higher than the electroweak scale;
\item
deviations from lepton number conservation are
calculable (finite).
\end{itemize}

Does the framework introduced by our assumption
yield a viable theory,
in view of
$m_R$ being much larger than the electroweak (ew.) scale? First we
remark that the renormalization-group evolution of the
coupling matrices $\Gamma_j$, $\Delta_j$ does not induce
flavour-changing elements, once we start with diagonal matrices; this
is due to the soft nature of the breaking
of the family lepton numbers. On the other hand,
there are strong experimental constraints \cite{groom}
on $\mu^- \to e^- \gamma$ and
on $\mu^- \to e^- e^+ e^-$. In our framework,
the branching ratio of the first
process is suppressed by $1/m_R^4$, the latter one by eight powers of
Yukawa couplings (the large scale enters only logarithmically in this
case) \cite{grimus}. Also the branching ratio for 
$Z \to e^- \mu^+ + e^+ \mu^-$ is suppressed by $1/m_R^4$ \cite{illana}.
Therefore,
breaking family lepton numbers \emph{softly} at a very high scale does
apparently lead
to a sensible theory
which is not in conflict with experimental results.

Within our framework we shall discuss in the forthcoming
sections three models,
realizing maximal atmospheric neutrino mixing, maximal solar
mixing, and bimaximal mixing \cite{bimaximal}, respectively.

\section{Maximal atmospheric neutrino mixing}
\label{maxatm}

It turns out that to enforce maximal atmospheric mixing we need
(at least) three scalar doublets ($n_H = 3$),
and we have to introduce two $Z_2$ symmetries:
\begin{eqnarray}
 & Z_2: &
\nu_{\mu R} \leftrightarrow \nu_{\tau R}\, , \;
D_\mu \leftrightarrow D_\tau\, , \;
\mu_R \leftrightarrow \tau_R\, , \;
\phi_3 \to -\phi_3\, ;
\label{Z2} \\
 & Z_2': &
\mu_R \to - \mu_R\, , \;
\tau_R \to - \tau_R\, , \;
\phi_2 \to - \phi_2\, , \;
\phi_3 \to -\phi_3\, .
\label{Z2'}
\end{eqnarray}
The left-handed lepton doublets are denoted by $D_\alpha$.
Fields not appearing in these equations transform trivially.
The motivation for $Z_2$ is quite straightforward. In $M_R$ this
symmetry leads to
\begin{equation}
\left( M_R \right)_{e \mu} = \left( M_R \right)_{e \tau}
\quad {\rm and} \quad
\left( M_R \right)_{\mu \mu} = \left( M_R \right)_{\tau \tau}\, .
\end{equation}
Because of $Z'_2$, the second and third scalar doublet have no Yukawa
couplings to the neutrino fields $\nu_R$. Therefore, the Dirac neutrino
mass matrix has the form
\begin{equation} \label{MD}
M_D = \mbox{diag} \left( a, b, b \right).
\end{equation}
As a consequence,
the light-neutrino Majorana mass matrix
has the same structure as $M_R$:
\begin{equation}\label{Mnu}
\mathcal{M}_\nu =
\left( \begin{array}{ccc}
x & y & y \\ y & z & w \\ y & w & z
\end{array} \right).
\end{equation}
Maximal atmospheric neutrino mixing and $U_{e3} = 0$
immediately follow from this structure of $\mathcal{M}_\nu$. This
structure of $\mathcal{M}_\nu$ in the basis where the charged-lepton mass
matrix is diagonal has previously been suggested by several authors
(e.g.\ Ref.~\cite{formMnu}).
We stress that in our case this structure results from a symmetry,
i.e., we have a \emph{model} and not just a \emph{texture} for
$\mathcal{M}_\nu$. In the phase convention used for $U$ of
Eq.~(\ref{U00}),
we obtain the neutrino mixing matrix
\begin{equation}\label{U}
U = \left( \begin{array}{ccc}
\cos \theta & \sin \theta & 0 \\
\sin \theta/\sqrt{2} & -\cos \theta/\sqrt{2} & -1/\sqrt{2} \\
\sin \theta/\sqrt{2} & -\cos \theta/\sqrt{2} &  1/\sqrt{2}
\end{array} \right) \,.
\end{equation}
One can check that the solar mixing angle $\theta$ is expressed by the
parameters of $\mathcal{M}_\nu$ of Eq.~(\ref{Mnu}) as
\begin{equation} \label{theta}
\tan 2\theta = 2\sqrt{2}\, \frac{|x^*y+y^*(z+w)|}{|z+w|^2-|x|^2}\, .
\end{equation}
By virtue of the symmetry $Z'_2$, the scalar doublets $\phi_{2,3}$
couple only to $\ell_R$. The non-zero VEV of
$\phi_3$ breaks $Z_2$ invariance in the charged-lepton sector already
at tree level and allows for $m_\mu \neq m_\tau$; in the
neutrino sector, where only $\phi_1$ couples,
and since $\phi_1$ transforms
trivially under $Z_2$, this symmetry is valid at the tree level.
For details see Ref.~\cite{grimus}.

It is instructive to recall the symmetries invoked.
The three $U(1)_{L_\alpha}$ ($\alpha = e,\mu,\tau$) are broken softly
at scale $m_R$ by $\mathcal{L}_M$; the symmetry $Z_2 \times Z'_2$ is
broken spontaneously, together with the SM gauge group, by the VEVs of
the three Higgs doublets.

In summary, the model of this section fulfills the following:
\begin{itemize}
\item
The atmospheric mixing angle $\psi$
is exactly $45^\circ$;
\item
$U_{e3}$ is exactly zero;
\item
the solar mixing angle $\theta$
is free, but large if one avoids finetunings;
\item
$\mms/\mma$ is free; it must be made small by finetuning;
\item
the seesaw mechanism is responsible
for the smallness of the neutrino masses $m_{1,2,3}$.
\end{itemize}
For the first and second items the rationale is given by the $Z_2$
symmetry together with the
softly broken family lepton numbers.

\section{Nearly maximal solar neutrino mixing}
\label{maxsol}

In this section we dispense with the discrete symmetries of the
previous section, and a single Higgs doublet is
sufficient. We take up again the idea of approximate
$\bar L = L_e - L_\mu - L_\tau$ invariance, this time implemented in
the $\mathcal{L}_M$ of Eq.~(\ref{LM}): we assume that there are two scales
$m_R \ll \bar m_R$ in $M_R$ such that $\bar L$ is conserved in
between.\footnote{An earlier variant of the idea of soft $\bar L$
breaking by $\mathcal{L}_M$ is found in
Ref.~\cite{Lbar+seesaw}.} Thus we have the following picture:
$$
\begin{array}{ccccccc}
\mbox{EW. SCALE} & \lll & m_R & \ll & \bar m_R & \lll &
\mbox{PLANCK (?) SCALE} \\
& & \uparrow & & \uparrow & & \uparrow \\
& & \bar L\ \mbox{broken} & & L_{e, \mu, \tau}\, \mbox{broken,}
& & L_{e,\mu,\tau} \\
& & & & \bar L\ \mbox{conserved} & & \mbox{conserved}
\end{array}
$$
\begin{figure}
\setlength{\unitlength}{1cm}
\begin{center}
\begin{picture}(8,5)
\put(2,0){\vector(0,1){5}}
\put(1.75,0.0){\line(1,0){0.5}}
\put(1.75,3.7){\line(1,0){0.5}}
\put(1.75,4.0){\line(1,0){0.5}}
\put(2.5,0.0){\makebox(0,0)[l]{$m_3 = 0$}}
\put(2.5,3.7){\makebox(0,0)[l]{$m_1$}}
\put(2.5,4.0){\makebox(0,0)[l]{$m_2$}}
\put(1,2.5){\makebox(0,0){(I)}}
\put(6,0){\vector(0,1){5}}
\put(5.75,0.0){\line(1,0){0.5}}
\put(5.75,1.2){\line(1,0){0.5}}
\put(5.75,1.5){\line(1,0){0.5}}
\put(5.75,4.0){\line(1,0){0.5}}
\put(6.5,1.2){\makebox(0,0)[l]{$m_1$}}
\put(6.5,1.5){\makebox(0,0)[l]{$m_2$}}
\put(6.5,4.0){\makebox(0,0)[l]{$m_3$}}
\put(5,2.5){\makebox(0,0){(II)}}
\end{picture}
\end{center}
\caption{Neutrino spectrum (I) shows the inverted hierarchy obtained
in the scenario discussed in Section \ref{Lbarinv}. For comparison,
spectrum (II) of the model of Section \ref{maxsol} is also
depicted. The latter spectrum, characterized by Eq.~(\ref{specII}),
is neither hierarchical nor of inverse
hierarchy. \label{nuspectra}}
\end{figure}
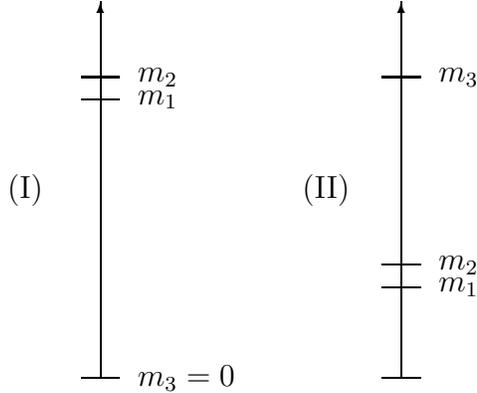
The individual lepton numbers are softly broken at $\bar m_R$, but the
linear combination $\bar L$ of the individual lepton numbers survives
down to $m_R$, where it is also softly broken.
Defining $\epsilon = m_R/\bar m_R \ll 1$, with our implementation of
approximate $\bar L$ invariance the matrix $M_R$ has the form
\begin{equation}\label{MRLbar}
M_R = \left( \begin{array}{ccc}
u & p / \epsilon & q / \epsilon \\
p / \epsilon & r & t \\
q / \epsilon & t & s
\end{array} \right),
\end{equation}
where $u$, $p$, $q$, $r$, $s$, and $t$ are assumed to be all of order
of magnitude $m_R$. Note that since $M_D$ is diagonal, $\bar L$ is not
violated in the Dirac neutrino mass matrix. After applying the seesaw
formula (\ref{seesaw}), with some lengthy algebra we obtain the
mass spectrum of the light neutrinos \cite{grimus}
\begin{equation}\label{spectrumII}
\begin{array}{ccc}
m_{1,2} & = & \epsilon m' \pm \epsilon^2 m'' \\
m_3 & = & m_0 + \epsilon m'_0
\end{array}
\quad \mbox{with} \quad m', \: m'', \: m_0, \: m'_0 \sim
(m_\nu^D)^2/m_R \,.
\end{equation}
Note that this mass spectrum is very different from the one obtained
in Section \ref{Lbarinv}, with approximate $\bar L$ invariance in
$\mathcal{M}_\nu$ itself, instead of in $M_R$. The two spectra are
compared in Figure \ref{nuspectra}. It is interesting that the
spectrum (\ref{spectrumII})
does not fit into either of the usual categories
``hierarchy'' and ``inverted hierarchy'',
since it has the properties
\begin{equation}\label{specII}
m_2 - m_1 \ll \frac{1}{2}(m_1 + m_2) \ll m_3 \,.
\end{equation}

In summary, the characteristic features of this model (for details see
Ref.~\cite{grimus}) are given by
\begin{equation}\label{features}
\mma \simeq m_0^2 \,, \quad \frac{\mms}{\mma} \sim \epsilon^3 \,,
\quad U_{e3} \sim \epsilon \,, \quad 1-\sin^2 2\theta \sim \epsilon^2
\,.
\end{equation}
While for the LOW solution of the solar neutrino deficit
we estimate, from the value of $\mms / \mma$,
$\epsilon \sim 1/30$, for the LMA MSW
solution we have $\epsilon \sim 1/4$. Thus, in the LMA MSW case the two
scales $m_R$ and $\bar m_R$ are rather close. Concentrating on the
LMA MSW solution and using the
order of magnitude estimate for $\theta$ in Eq.~(\ref{features})
despite of the not very small $\epsilon$, we
obtain $\theta \sim 38^\circ$. This is not so far from the best-fit
value for the solar mixing
angle\footnote{This is the value of the fit with a free $^8$B flux;
the value of the fit where the $^8$B flux is restricted by the Solar
Standard Model is slightly lower \cite{krastev}.}
of Ref.~\cite{krastev},
$\theta \simeq 32^\circ$. Furthermore, $|U_{e3}|^2 \sim \epsilon^2$
might be on the border of
becoming discernible in the data.
We want to stress,
however, that we have presented here
only a crude picture of the situation,
since ratios of the undetermined constants in $M_R$ of Eq.~(\ref{MRLbar}),
which are assumed to be all of similar order of magnitude,
enter into the precise version of Eq.~(\ref{features}) (see
Ref.~\cite{grimus}). Concerning the atmospheric mixing angle $\psi$, we
can only say that it must be large but we have no rationale for
setting it very close to $45^0$.

As we have seen, in the scheme of soft $\bar L$ breaking developed here, the
solar mixing angle is not necessarily close to $45^\circ$, and $U_{e3}$
can deviate appreciably from zero. Thus, not only in the neutrino mass
spectra but also in neutrino mixing, we have striking differences
between the scenario of this section and the one
presented in Section \ref{Lbarinv}.

\section{Bimaximal neutrino mixing}
\label{bimax}

Bimaximal neutrino mixing can be obtained by combining the models of
Sections \ref{maxatm} and \ref{maxsol}.
Then, the
number of Higgs doublets must again be three,
and we must again have the $Z_2 \times Z'_2$ symmetry
in the Lagrangian prior to spontaneous symmetry
breaking with, moreover, intermediate $\bar L$ invariance in $M_R$.
In this scenario we
preserve the good relations
$1 - \sin^2 2\theta \sim \epsilon^2$ and $\mms/\mma \sim \epsilon^3$,
and are additionally rewarded with $\psi = 45^\circ$ and $U_{e3} = 0$
at tree level.

\section{Conclusions}

In this conference report we have first discussed the scenario with
approximate $\bar L = L_e - L_\mu - L_\tau$ invariance in the mass
matrix of the light neutrinos \cite{barbieri}.
We have then
considered a novel framework for imposing maximal neutrino mixing,
either for atmospheric neutrinos or for solar neutrinos or for both,
which was introduced in Ref.~\cite{grimus}. This framework consists of
the SM with three families and three right-handed neutrino singlets
with a heavy
Majorana mass term with mass matrix $M_R$, such that
the three individual lepton numbers $L_{e,\mu,\tau}$ are
conserved in the Yukawa couplings but \emph{softly broken} by the
Majorana mass term  of the right-handed singlets.  This is a very
interesting scenario where deviations from $L_{e,\mu,\tau}$ and
from $L = L_e + L_\mu + L_\tau$ conservation are calculable and
controlled by the mass matrix $M_R$. The mass matrix of the charged
leptons is automatically diagonal.

Within this framework we have imposed different symmetries in order to
obtain three different models of neutrino mixing:
\begin{enumerate}
\item
$Z_2 \times Z'_2$ symmetry: In this case, where three Higgs doublets are
necessary, we have obtained an
atmospheric mixing angle of $45^\circ$, $U_{e3} = 0$, and a large but
otherwise free solar mixing angle. We want to stress again that
{\em symmetries} are responsible for these results,
not ``textures'' or \emph{Ans\"atze}.
\item \label{II}
Intermediate $\bar L$ conservation in $M_R$: Here we need only one
Higgs doublet, but we have two scales
in $M_R$ such that at the higher scale the individual family lepton
numbers are broken down softly to the linear combination $\bar L$,
which is then broken, also softly, at the lower scale. In this case the
atmospheric mixing angle is free but large in general,
whereas the solar mixing angle should be
in the range around $44^\circ$, with the LOW solution of the solar
neutrino deficit, 
or around $38^\circ$, with the LMA MSW solution.
Also $U_{e3}$ can differ considerably from zero.
\item Combination of the two symmetries
leads to bimaximal mixing with $U_{e3}$ again vanishing.
\end{enumerate}

We have compared the scenario where $\bar L$ invariance is
imposed directly on the light-neutrino mass matrix
(Section \ref{Lbarinv}) with
the new scheme of item \ref{II}
where $\bar L$ invariance is rather imposed
on $M_R$ (Section \ref{maxsol}).

Our new mechanisms for enforcing maximal neutrino mixings are
accommodated in a rather simple extension of the SM. In particular, neither
supersymmetry nor Grand Unified Theories are invoked. Quite on the
contrary, it seems difficult to incorporate our framework into
a GUT.

\section*{Acknowledgement}
W.G. thanks the organizers of the school for their warm hospitality.

\end{document}